\documentclass[graybox]{svmult}

\usepackage{mathptmx}       
\usepackage{helvet}         
\usepackage{courier}        
\usepackage{type1cm}        
\usepackage{makeidx}         
\usepackage{graphicx}        
\usepackage{multicol}        
\usepackage[bottom]{footmisc}

\makeindex             

\begin{document}

\title*{Can Deep Altruism Sustain Space Settlement?}
\author{Jacob Haqq-Misra}
\institute{Jacob Haqq-Misra \at Blue Marble Space Institute of Science, 1001 4th Ave, Suite 3201, Seattle, Washington 98154 USA, \email{jacob@bmsis.org}\\
Published in \emph{The Human Factor in a Mission to Mars: An Interdisciplinary Approach}, K. Szocik (Ed.), Springer.}
%
%
\maketitle

\abstract*{Space settlement represents a long-term human effort that requires
unprecedented coordination across successive generations. In this chapter, 
I develop a comparative hierarchy for the value of long-term projects based upon their 
benefits to culture, their development of infrastructure, and their
contributions to lasting information. I next draw upon the
concept of the time capsule as an analogy, which enables a comparison of
historical examples of projects across generational, intergenerational, and
deep time. The concept of deep altruism can then be defined 
as the selfless pursuit of informational
value for the well-being of others in the distant future.  The first steps toward supporting an effort like space settlement
through deep altruism would establish governance and funding models that begin
to support ambitions with intergenerational succession.}

\abstract{Space settlement represents a long-term human effort that requires
unprecedented coordination across successive generations. In this chapter, 
I develop a comparative hierarchy for the value of long-term projects based upon their 
benefits to culture, their development of infrastructure, and their
contributions to lasting information. I next draw upon the
concept of the time capsule as an analogy, which enables a comparison of
historical examples of projects across generational, intergenerational, and
deep time. The concept of deep altruism can then be defined 
as the selfless pursuit of informational
value for the well-being of others in the distant future.  The first steps toward supporting an effort like space settlement
through deep altruism would establish governance and funding models that begin
to support ambitions with intergenerational succession.}

\section{Introduction}
\label{sec:intro}

The prospect of human settlements on Mars 
is becoming increasingly technologically feasible. Efforts by SpaceX, Deep Space Industries, Planetary Resources, and
other private space corporations now fall in rank with government space agencies such as
NASA, ESA (European Space Agency), JAXA (Japan Aerospace Exploration Agency), IRSO (Indian Space Research Organization),
RFSA (Russian Federal Space Agengy, or Roscomos), and CNSA (China National Space Administration). 
Many of these corporate and government entities
are developing successive plans to visit asteroids or Mars within the next few decades 
\cite{seedhouse2010,nasa2015}, which are
beginning to show prospects for economic gain in addition to scientific return. These recent
developments all suggest that Elon Musk's vision of human civilization becoming a ``multiplanetary
species'' \cite{musk2017} could be realized in the coming centuries.

Technological advances that enable humans to settle on another planet or extract resources
from planetary bodies must be matched by parallel advances in civilizational ethics. A
lack of moral progress risks the danger of perpetuating the problem of the commons and other
harmful colonial attitudes as human civilization ventures into space. Land use policies on Mars must 
account for the finite extent of resources while also respecting the non-appropriation
principle of the Outer Space Treaty that restricts national claims to sovereignty \cite{haqqmisra2015}.
New governance models, either by modification of the Outer Space Treaty or by the creation of 
new international institutions, can promote solutions to the equitable sharing of Mars
and other space-based resources \cite{ehrenfreund2013,cockell2015,haqqmisra2016,bruhns2016}; however, implementation of any such idea would require
commitment and cooperation from most of the major space-faring nations as a minimum. Nevertheless, 
establishing sustainable space settlements will require that humanity begin advancing 
its ethics in tandem with technology, prior to the arrival of the first humans on Mars.

Innovative approaches to fundraising, charitable giving, and other means of financing can further enable long-term and intergenerational initiatives such as space settlement.
Corporations like IBM, Lloyd's of London, and the Swedish National Bank all hold experience 
in maintaining business tradition and financing ventures over decadal timescales and longer, while organizations
like the Rockefeller Foundation and the Carnegie Corporation have promoted the advancement of knowledge
through charitable contributions for more than a century. 
Crowdsourced models could also provide sustained funding for space
settlement or other long-term scientific initiatives, such as the search for extraterrestrial intelligence (SETI).
Specialized financial products could be tailored toward individual scientific
objectives or development goals, such as a ‘lottery bond’ debt security, which could provide a regular stream of income to ambitious projects while also providing consumers with a direct return on investment \cite{haqqmisra2018}.
The success of commercially-driven space settlement will inevitably require tremendous
financial foresight that could benefit from a donor, group, or crowd willing to invest in the distant future of
humanity.

In this chapter, I define the concept of ``deep altruism'' as ambitious human efforts with  
high informational value across millennial timescales that do not provide direct benefits to the initial benefactor.
I begin by drawing upon the analogy of a time capsule, which represents an abstraction of a long-term
human effort with the intention of providing value to future generations.
I then define a relative value scale based upon cultural, structural, and informational
motivations. I next consider historical and contemporary examples of the completion time
for projects across generational, intergenerational, and deep time. 
I conclude by discussing the feasibility of establishing a space
settlement based upon a deep altruistic funding model.

\section{Time Capsules}

The time capsule serves as an example of a long-term effort with altruistic intentions, which highlights some of the 
unique challenges in approaching altruism across deep time. The Oxford English Dictionary defines ``time capsule'' 
as ``a container used to store for posterity a selection of objects thought to be representative of a particular moment in time.'' Time capsules represent an attempt at preserving value from a particular time in history for the benefit of 
other people in the future.

The concept of a time capsule today usually refers to an object constructed with intention, in order to purposefully 
commemorate a particular time by preserving the value of its memory. Archaeological discoveries
also provide scholars today with new information, but such discoveries rarely encounter concerted efforts 
of cultures from the past attempting to communicate with us today. The first deliberately constructed time capsule, 
the ``Century Safe,'' was featured at the 1876 World's Fair in Philadelphia, which established
the modern tradition of intentional time capsules.

Time capsules can further be distinguished by the intended future audience, which corresponds to the length 
of time that the capsule must remain preserved. ``Target-dated'' time capsules specify a particular length of time
(\textit{e.g.,} a century) when the capsule will eventually be opened; ``deliberately infinite'' time capsules
are preserved in perpetuity until future generations eventually decide to open the capsule \cite{jarvis2002}. 
Both types of capsules can suffer from the lack of complete information, such as including toys or trinkets
without explanation of their significance, which limits the value of the objects to historians.
An ideal time capsule constructed today, to maximize future value, should include a ``full set of cultural-technical information drawn from the whole of human world culture'' with a ``10,000-year target span date'' \cite{jarvis2002}.
Other features for an effective time capsule include the existence of redundant copies to protect
against loss, as well as electronic access to the contents in order to increase transparency and maintain interest.

A time capsule with a target date that approaches millennial scales or longer serves as an illustration
of a general altruistic effort that operates over deep time. Short-term time capsules with a target date on 
the scale of decades could provide direct benefit to people who were living when the capsule was sealed.
A longer-term time capsule with a target date on the scale of a century would not necessarily benefit
the individuals who sealed the capsule, but their grandchildren 
or great-grandchildren would likely benefit from the value of opening the capsule. A time capsule with a target 
date on a deep time scale of millennia or longer would represent a genuine altruistic effort, as the individuals 
or community that sealed the capsule would likely have no direct connection to the people opening the capsule.
A time capsule motivated by a sense of deep altruism would seek to provide relevant and contextual
information to humanity's distant descendants. 

The problem of preserving the contents and knowledge of a time capsule over millennia poses similar challenges 
to other long-term projects attempted across history. The settlement of space, including the development of 
permanent human habitats and even the terraforming of an entire planet, will require unprecedented 
cooperation and coordination across deep time. As with the time capsule, any long-term
effort like space settlement must maintain informational relevance and sustain operations across generations
in order for its future value to be realized.

\section{Cultural, Structural, and Informational Value}
\label{sec:value}

\begin{figure}
\centering
\includegraphics[scale=.61]{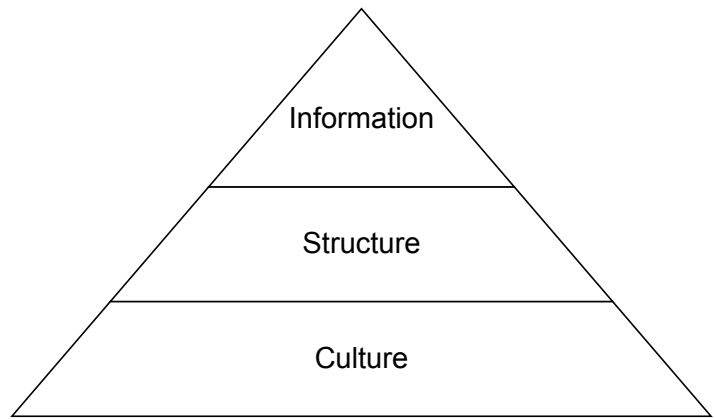}
\caption{The hierarchical relationship between cultural, structural, and informational value provides a 
framework for comparing the relative value of long-term human projects.}
\label{fig:1}
\end{figure}

In order to further unpack the concept of deep altruism, I define a framework for comparing
the relative benefits derived from long-term human projects. 
The investment cost in a long-term project is not necessarily a reliable indicator of its realized value, 
as seemingly inexpensive items may show themselves to be priceless when retrieved by its future recipients. 
Apparent success of a project today is also not a reliable indicator of future value, as human civilization
will likely evolve its infrastructure and preferences across the next millennium. 
Instead, I draw upon value theory to establish a comparative scale for long-term projects 
based upon the ultimate type of value realized by future generations. Value theory is a broad approach 
in ethics for defining relative degrees of goodness, benefits, or other desirable features, 
which enables comparison of the relative value of objects or actions even if further quantification
is difficult. For this analysis, I develop a specific hierarchical approach toward valuation of 
long-term projects based upon cultural, structural, and informational value.
This value hierarchy is summarized as a pyramid chart in Fig. \ref{fig:1}, with culture
at the base, structure in the middle, and information at the apex. 

The first level of cultural value refers to a long-term project that is primarily concerned with aesthetic factors, 
the preservation of tradition, and other features central to the group's identity. A project motivated
by cultural value would seek to enable future descendents to experience similar, 
or greater, appreciation of the originating culture. Cultural value resides at the foundation of the value 
hierarchy because culture is inherent to all human activities. Thus, the remaining levels in the value 
hierarchy necessarily include culture in addition to other value considerations.

Structural value is the second level, which describes a long-term project that is primarily concerned with the 
preservation of materials, buildings, or other engineering feats. Structurally motivated projects 
require diligent maintenance in order to preserve the integrity of the original 
construction for future descendants. Many structures are intended primarily for utilitarian purposes, 
such as shelter or storage; the long-term preservation of such structures can be accomplished 
by active efforts by governments, individuals, or private organizations. However, some structures are more strongly linked with culture, such as religious temples, burial sites, or ancestral shrines. 
The value hierarchy therefore includes an intermediate level on the culture/structure boundary
to account for structures that are maintained by time-tested cultural traditions.

The final level of informational value represents the realization of culture and structure 
to provide long-lasting benefits to human knowledge. A project motivated by informational value
would seek to build upon knowledge traditions to enable future generations to solve major problems.
Informational value provides tools and methods that enable solutions, whereas cultural value 
only provides a means of preserving information. Some projects involve significant engineering and management
innovation in order to support the acquisition of knowledge, as is increasingly required by 
``big science'' projects in physics, astronomy, and materials science. The value hierarchy
therefore includes an intermediate level at the structural/informational boundary to account for 
physical structures that support efforts at achieving new knowledge.

This three-tiered hierarchy of cultural, structural, and informational value provides a relative scale 
for ranking long-term human projects across history. The purpose of such a relative value scale
is not to judge the historical merits of any of these efforts; instead, this approach enables 
an analysis of the factors that enable the successful preservation of each type of 
value across deep time.

\section{Completion Time}
\label{sec:time}

History is abundant with examples of long-term projects conducted across generational, intergenerational,
and deep time scales. The completion time for such projects is defined as the amount of time
between the initial conception of the idea and its final execution, analogous to the duration of a time capsule. 
Some projects include a target date for completion 
after a finite amount of writing, construction, or analysis---similar to a target-dated time capsule.
Other projects are deliberately maintained in perpetuity, with no intention to cease operations,
comparable to a deliberately infinite time capsule.

\begin{figure}
\centering
\includegraphics[scale=.61]{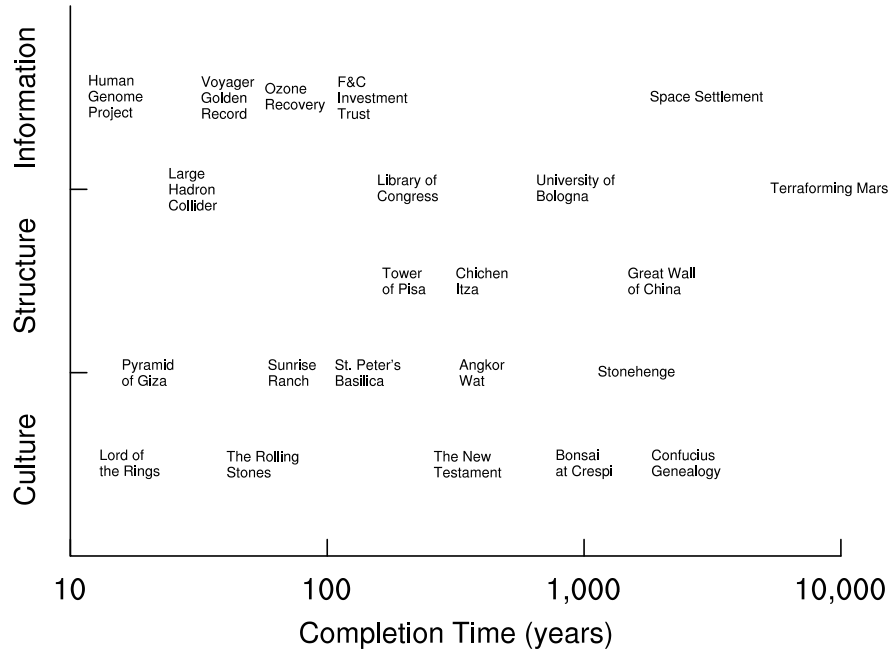}
\caption{Examples of long-term human projects plotted as a function of total completion time versus value, where the value is categorized as cultural, structural, or informational.}
\label{fig:2}
\end{figure}

A range of historical and contemporary examples of long-term projects is shown in 
Fig. \ref{fig:2}, with completion time on the horizontal axis and value on the vertical axis.
Fig. \ref{fig:2} is intended to be an illustrative, rather than comprehensive, collection 
of projects conducted over 
generational, intergenerational, and deep time with cultural, structural, and informational
value motivations. 
This visualization of successful long-term projects enables a comparison 
of the factors required for sustaining altruistic activity over deep time.

\subsection{Generational Time}

Projects that occur within a single generation have a completion time between about 10 to 100 years. 
Generational time efforts often have a target date for completion, usually with the intention 
of finishing within the lifetime of the project's originator. The \textit{Lord of the Rings} trilogy
by J. R. R. Tolkien was written in stages over a period of 12 years, 
which represents a decadal-scale work of cultural value that still persists in completed form today. 
The Great Pyramid of Giza is an example of a work of structural value, which took about 20
years to build. Informational projects can also have a finite duration, such as the 30-year construction 
time to build the Large Hadron Collider, the world's most powerful particle accelerator. 
Likewise, the Human Genome Project 
required 13 years of collaborative research to obtain a complete mapping of all human genes.
Successful generational time projects with a finite target date usually 
allow the originator to personally experience the full value upon completion, even if the final
product persists for much longer.

Some generational projects are deliberately infinite by design and could extend into longer intergenerational timescales. 
The Rolling Stones is the longest continuously-performing rock band, founded in 1962 and still performing
with some of its original members; conceivably, the franchise could persist beyond the lifetime of the individual
band members. One of the longest-running intentional communities in the US is Sunrise Ranch, operated 
by the Emissaries of Divine Light since 1945.
The golden records on board both Voyager spacecraft, launched in 1977, 
contains encoded knowledge about Earth and its inhabitants that will continue drifting
through interstellar space indefinitely. 
The ozone hole is slowly recovering as the human use of chlorofluorocarbons (CFCs) has declined, 
but projections indicate 70 years or longer before the Antarctic ozone layer recovers to pre-1980 levels.
Deliberately infinite projects necessarily begin at the generational scale, but if they are successful
then they will continue beyond the lifetime of the originator to be tended by the next generation.

\subsection{Intergenerational Time}

The completion time for intergenerational projects is about 100 to 1,000 years. Intergenerational time 
efforts may have a goal or objective but not necessarily a reliable target date for completion.
The time between the writing and canonization of the New Testament was about 350 years; 
although the text has been faithfully preserved, the original authors of most of the New Testament remain unknown.
Many cathedrals, such as St. Peter's Bassilica in the Vatican, took a century or longer to construct, 
while the Angkor Wat temple complex in Cambodia was built in stages over a period of 400 years. 
The Tower of Pisa took 199 years to build in three stages, showing evidence of sinking with the completion of 
each floor. The Chichen Itza building complex of the Maya took about 400 years to construct, which remains 
a draw for tourists today. Many intergenerational structures and cultural artifacts still provide 
value for people today, even if for different purposes than intended.

Other intergenerational projects have maintained deliberately infinite operations, with no 
defined point of completion. The Library of Congress is one example of a robust infrastructure 
that has sustained the acquisition and cataloging of human knowledge for over 200 years. 
with the 150-year old Foreign \& Colonial Investment Trust holding the record for being
the longest-running investment fund. Educational institutions such as the University of Bologna
and Oxford University have managed to sustain their structures and operations for nearly a millennium.
Institutions that manage to maintain deliberately 
infinite operations across multiple generations, while successfully adapting to change, 
will eventually approach the threshold of deep time.

\subsection{Deep Time}

A project that operates within deep time has a completion time of 1,000 years or longer.
Deep time efforts must contend with dramatic shifts in geopolitics, changes in the Earth system, 
and other factors that remain less volatile at shorter timescales. 
Structures such as Stonehenge were constructed in stages over about 1,500 years, perhaps with a general
goal but likely no target completion date. The Great Wall of China was built, extended, and repaired 
by multiple dynasties over about 2,000 years, with the goal of protecting the northern border from invading armies.
Finite duration projects that operate over deep time represent long-lasting organizations that 
were able to maintain culture and engineering tradition over the course of drastic historical changes.

Deliberately infinite deep time projects require careful attention by successive generations in 
order to preserve knowledge that would otherwise be forgotten. Unlike structural value, which can persist 
even if the founding culture becomes extinct, cultural value can be lost if not preserved
through written and oral tradition. The Ficus Bonsai Tree at Crespi, Italy, is among the oldest
in the world and has received daily care for the past 1,000 years. Similarly, the 
genealogy of Confucius has been dutifully preserved for 2,500 years, enduring through numerous dynasties
and political revolutions. 
Such deliberately infinite efforts at preserving culture seem likely to continue for the 
foreseeable future, as long as people continue to recognize their value.

Few, if any, human ambitions have successfully managed to realize informational value across deep time. 
Contemporary efforts to enable the human settlement of space represent a deep time ambition to achieve informational 
value. Space settlement is an infinite duration project that will likely be developed in stages,
with an idealistic goal of enabling self-sustaining human populations that no longer require support from Earth. 
The timescale for achieving an initial human presence in space may be generational, but any long-lasting
space settlement must successfully traverse deep time in order to demonstrate its autonomy.
Even more audacious ideas to terraform a planet like Mars, so that it could sustain plant life 
and a breathable atmosphere, represent a deep time effort with a finite target date; however, 
transforming an entire planetary system would take 10,000 years or longer of patient 
monitoring in order to reach the desired climate state. If any such plans for the 
permanent human settlement of space actually do begin to take shape, then they will represent the first
intentional effort at pursuing informational value across deep time.

\section{Deep Altruism}

I can now revisit the definition of deep altruism, drawing upon the relative value scale for long-term 
projects and the discussion of completion time.
The word ``altruism'' is defined by the Oxford English Dictionary as ``disinterested or selfless concern for the well-being of others,
especially as a principle of action.'' In the context of long-term projects over deep time, altruism refers to selfless
concern for the well-being of others in the distant future. The preservation of cultural tradition can include
altruistic elements, but such efforts do not necessarily foster new methods of systematically increasing total well-being. Conversely, altruistic pursuit of informational value can expand knowledge and enable solutions
to problems that significantly improve the quality of life. Deep altruism can therefore be defined as shown in the box below.

\begin{svgraybox}
\textbf{Deep altruism} is the selfless pursuit of informational value for the well-being of others in the distant future.
\end{svgraybox}

The assortment of projects in Fig. \ref{fig:2} show a lack of examples with informational value and millennial
completion time, although such an approach is necessary for the daunting task of settling space. 
However, many projects beginning 
today are motivated by altruistic intentions, with target dates that approach deep timescales.
The Clock of the Long Now is being constructed with the intention of keeping time for 10,000 years; the clock
is being funded by Bezos Expeditions and currently resides on land owned by Jeff Bezos. 
The Letters of Utrecht is a collective poem carved into cobblestones along the city's streets,
with a new letter added every Saturday. The Letters of Utrecht began in 2012 and is intended to continue
in perpetuity---or as long as the citizens of Utrecht permit.
A record-setting organ performance of ``As Slow as Possible'' composed by John Cage is underway 
at the St. Burchardi church in Germany, with the first note beginning in 2001 and the piece 
reaching a finale in 2640.
Although all of these efforts presently have been able to secure enough resources to maintain operations,
they are still in the initial generational phase where the project originators are still alive and involved.
The long-term success of these and other altruistic efforts will require effective succession between generations
in order to transition into intergenerational and deep time.

Why would an individual or organization choose to engage in deep altruism?
Reciprocal altruism features in many instances of biology, as natural selection pressures
can operate against individuals who choose selfish or cheater behavior in cooperative groups \cite{trivers1971}.
Non-reciprocal altruism seems to be a unique feature of humans (and possibly a few other primate species),
with a less obvious explanation for the evolutionary benefits of selfless concern for even complete strangers.
Understanding non-reciprocal altruism remains an ongoing area of research, with some analyses suggesting that 
social distance can correlate to expectations of reciprocal altruism \cite{brinkers1999,takahashi2007}. Others find that
a fraction of a population may be inclined to act with non-reciprocal altruism, even if the majority
chooses otherwise \cite{johannesson2000}.
Deep altruism similarly features an extreme degree of non-reciprocity, with no direct benefits to the originator---who 
may even be long forgotten by the future recipients of the completed effort. 
Musk and Bezos represent wealthy individuals who aspire to leverage their resources toward bold 
ambitions that they will not personally see to conclusion. Perhaps they hope to secure their names in history
through such grand investments, but possibly they are also motivated by the desire to 
improve human civilization's capabilities of intentionally cooperating across intergenerational and deep time. 
A benefactor acting out a sense of deep altruism requires a vision for the species that extends beyond their 
own life; such a person may be motivated by the desire to alleviate suffering in the world, to 
increase the sustainability of civilization, or other global objectives that remain beyond the 
capability of any individual or generation to solve. 
Altruism in general, and deep altruism in particular, may be a uniquely human response to the problems 
elicited by civilization itself. Selfless action for the well-being of others across deep time
might be one of the only available approaches for building a better long-term future for humanity.

\section{Conclusion}

Deep altruism remains a viable option for supporting human ambitions to settle space, as long
as the initial benefactor can effectively transition the management, leadership, and vision 
of the effort to subsequent generations. Deep altruistic projects have a duration that is effectively
infinite, as the succession of operations across generations is more critical to success than estimating a completion date.
Any effort based upon a deep altruistic model must effectively communicate the vision that inspired the 
founder, and likewise must establish a secure source of funds that can also persist across generations. 
Wealthy individuals and institutions could conceivably finance such ventures, as long as they consider the benefits 
to the distant future as an adequate justification for investment today.

From the perspective of space settlement, which necessarily must operate with deep time in mind, 
the antithesis to deep altruism can be approximated as ``deep egoism.'' A mindset fostered
by deep egoism would assess the value of long-term investment by its propensity to 
benefit self, kin, colleagues, and descendants over others. The calculated return on investment
for some asteroid mining ventures is predicted to exceed trillion of dollars; although these profits
may not be realized by today's investors in asteroid mining technology, this conventional
funding model would shunt the resulting wealth to the hands of individuals or organizations intended
by the first investors. A long-term effort at extracting space resources for the purpose of building
corporate or government wealth could be another approach to improve humanity's ability to operate over long 
timescales. 
It is important to note that deep egoism could conceivably sustain long-term efforts such as space settlement. 
For-profit entities face their own internal or external pressures to continue sustaining operations as long as 
they remain profitable; such pressures can drive innovation and allow companies to adapt to changing market pressures
and new technology. Deep egoism could even motivate efforts across deep time and will likely be a significant driver of the
near-term human exploration of space. Deep egoism resonates more strongly with modern capitalist ideals, although such 
an approach would risk failure if it is unable to continually provide a return on investment.

Humanity can take steps toward enabling altruistic efforts over deep time scales.
Government and private granting agencies could develop competitive funding programs with
decadal and longer performance periods in order to promote intergenerational pursuits
of informational value.
Wealthy individuals and organizations also hold the resources to establish their own 
long-term efforts by planning for intergenerational succession from the start.
Crowdsourcing provides yet another approach for the general public to engage in supporting long-term efforts, 
with tools such as online crowdfunding and distributed consensus decision-making models providing 
the foundations for crowd-driven deep altruism. 
Striving toward informational value across intergenerational timescales will 
pave the way for extending such pursuits into deep time.

Space settlement will unfold in a piecewise manner, likely by a combination of government and private 
actors with a range of motivations. Commercial interests remain an important 
driver of the near-term space economy and could be a significant factor in developing 
the physical infrastructure required for space settlement. But deep egoism alone 
may be insufficient to sustain space settlement across millenia. 
If humanity genuinely intends to develop permanent settlements on Mars and in space, then
it will inevitably be forced to develop new institutional governance models driven by deep altruism .

\begin{acknowledgement}
I thank Lauren Seyler, Sanjoy Som, and Martin Elvis for helpful suggestions. This research did not receive any specific grant from funding agencies in the public, commercial, or not-for-profit sectors.
\end{acknowledgement}

%
%
%
%
%

\end{document}